# Demonstration of traveling-wave interactions between spontaneous photon emissions and atoms in a chiral F–P cavity

Jiajin Lu, [1,2] Minjie Wang, [1,2] Haole Jiao, [1,2] Xiang Chen, [1,2] Hongze Zhang, [1,2] Shujing Li [1,2,†] and Hai Wang [1,2,†]

[1]*State Key Laboratory of Quantum Optics and Quantum Optics Devices, Institute of Opto-Electronics, Shanxi University, Taiyuan 030006, China*

[2]*Collaborative Innovation Center of Extreme Optics, Shanxi University, Taiyuan 0300*

**ABSTRACT.** The enhancement of atom–photon interactions with F–P cavities provides a suitable platform for studying quantum optics and atomic physics. However, the emission fields in linear F–P cavities are in the standing-wave mode, which leads to non-uniform atom–photon coupling and a short storage lifetime of cavity-enhanced spin-wave quantum storages. This study experimentally demonstrates traveling-wave atom–light interactions in an F–P cavity that can preserve light helicity. First, a bias magnetic field is applied along the z-axis to define the quantization axis, which lifts the Zeeman degeneracy and breaks the time reversal symmetry. Next, non-classically correlated pairs of Stokes photons and spin waves are produced based on the Duan–Lukin–Cirac–Zoller scheme. The Stokes photons initially emitted from a single circularly polarized ( $\sigma^-$ ) atomic transition have left- and right-hand circular polarizations when propagating along the +z (forward) and –z (backward) directions, which are preserved in the chiral cavity within the atom–photon interaction region. Thus, when the forward and backward Stokes fields resonate with the cavity, they may interact with the atoms in a traveling-wave manner. This is confirmed by measuring the time-dependent retrieval efficiencies of spin waves correlated with the forward and backward Stokes fields. This work paves the way for demonstrating traveling-wave atom–photon interactions in F–P cavities.

## I. INTRODUCTION.

It is well known that couplings (interactions) between cold and ultracold atoms with light fields can be significantly enhanced by cavities. Over the past two decades, atom–cavity systems have been vital tools for investigating atom–photon quantum interfaces [1-3] and quantum phase transitions in many-body physics [4]. By trapping individual atoms and coupling them to high-finesse cavities, researchers have demonstrated cavity quantum electrodynamics, where atomic spontaneous emission is effectively constrained in the cavity axis, making the system an effective quantum node [3]. With cold atomic ensembles inserted into optical cavities, high-storage-efficiency Duan–Lukin–Cirac–Zoller (DLCZ) quantum memories and electromagnetically induced transparency quantum storage have been demonstrated [2, 5, 6]. The enhanced collective coupling of Rydberg atomic ensembles with light fields via ring cavities, which enables quantum logic operations on photon pulses, has been experimentally demonstrated [7-10]. With the loading of quantum gases, that is, ultracold atoms (BEC), into high-finesse cavities, many-body physics has been theoretically and experimentally studied [11-27].

On the other hand, nonreciprocity couplings (NRCs) between light and quantum emitters, which refer to the unidirectional emission, scattering, and absorption of photons by quantum emitters, have been widely studied [28]. Experiments on NRCs have been conducted using quantum emitters in optical nanofibers [29] and photonic crystal waveguides [30]. In these nanophotonic structures, the optical fields are tightly confined in the transverse direction, inducing spin–momentum locking, which means that the polarization (spin) of the light field is locked to its propagation direction [31, 32]. Spin–momentum locking enables light emission from circularly polarized (chiral) dipole transitions to be direction-dependent.

In free-space systems, the light emission from circularly polarized transitions of atoms has a well-defined polarization along the z-axis, defined by a bias magnetic field. On this basis, Simon et al. experimentally demonstrated the isolation of the running-wave mode of a moderate-finesse optical cavity by breaking the time reversal symmetry (TRS) [33], and Yelin et al. studied chirality-induced emergent spin–orbit coupling [34], chirality-dependent photon transport, and helical superradiance [35]. Recently, the handedness–momentum correlation of spontaneous Raman scattering on a circularly polarized dipole transition was observed with a cold atomic ensemble in a ring cavity [36]. Thus far, NRCs between atoms and light fields in standing-wave F–P cavities have remained exclusive. F–P cavities have simpler structures than ring cavities and are widely used for studying quantum optics [1-3]. Recently, fiber-based F–P cavities have been explored [37, 38], and their coupling with atoms has been experimentally demonstrated [39]. However, in the standard F–P cavity, both forward- and backward-direction propagating fields are superposed, which enables the spatial modulation of atom–photon interactions. Standing-wave atom–light interactions are undesirable in many quantum optics experiments. For example, for the spin-wave storage in cold atomic ensembles coupled to an F–P cavity, the superposition of the forward and backward write-out (Stokes) photons leads to a short-lived spin-wave storage [2]. To overcome such a short lifetime, Bao et al. used a ring cavity to distinguish backward-scattered from forward-scattered Stokes photons and obtained long-lived spin-wave storage [5]. However, ring cavities have significant drawbacks compared with the standing-wave cavity: (1) complex cavity configuration, which is not suitable for large-size multiplexed atom–photon interfaces; (2) longer cavity lengths, which require longer optical pulse widths to match the long cavity, leading to a slower rate of quantum information processing. In addition, many quantum information processing schemes using atom–photon interactions in cavities require spatially uniform coupling [40-42].

Building on these considerations, this study realizes a traveling-wave interaction between $^{87}$Rb atoms and spontaneous Raman light emission in a chiral F–P cavity. Chiral F–P cavities using metamaterial mirrors [43] have been used for enantiomer discrimination [44-46]. In contrast to these studies, the proposed chiral F–P cavity is formed by inserting two quarter-wave plates on either side of the atomic ensembles, which preserves the light helicity in the atom–cavity interaction region. The presented experiment was demonstrated via the DLCZ scheme [47] in cold $^{87}$Rb atoms [48], where the atoms were prepared in the desired Zeeman state via optical pumping. "Write" pulses are applied onto the cold atomic ensemble to generate Stokes photons that are non-classically correlated with spin waves, where the excitation probability is $\chi \approx 1\%$ . The chiral F–P cavity resonates with the Stokes photons emitted from a circularly-polarized ($\sigma^-$) transition of the atoms. The interactions of Stokes photons with atoms are nonreciprocal in the chiral F–P cavity.

## II. EXPERIMENTAL SETUP.

Fig. 1a presents the experimental setup. The atoms are a cloud of cold $^{87}$Rb atoms released from a magneto-optical trap (MOT). They are coupled to an F–P cavity comprised of $M_1$ and $M_2$ mirrors. Two ($\lambda/4$) plates, labeled $QW_L$ and $QW_R$, are inserted in the cavity, where $QW_L$ is placed on the left side and $QW_R$ on the right side of the atomic cloud. A beam displacer (BD) is also inserted in the cavity to control whether the forward or backward Stokes fields are resonant with the cavity (cf. below). Fig. 1b (1c) shows the relative atomic transitions of $^{87}$Rb atoms that interact with the "write" laser and the emissions of Stokes photons (the "read" laser and the emissions of anti-Stokes photons), where a bias magnetic field $B_0 = 10$ G is applied along the z-axis to define the quantization axis, thereby lifting the degeneracy of Zeeman sublevels and breaking the TRS. Fig. 1e provides an intuitive diagram for understanding the chiral atom–light interactions. The transition $|s\rangle = |5S_{1/2}, F=2, m_F=1\rangle \leftrightarrow |e\rangle = |5P_{1/2}, F'=1, m_F=0\rangle$ ($|s'\rangle = |5S_{1/2}, F=2, m_F=-1\rangle \leftrightarrow |e\rangle = |5P_{1/2}, F'=1, m_F=0\rangle$) is a $\sigma^-$- ($\sigma^+$-) circularly polarized transition [49]. The light emission from the $\sigma^-$- ($\sigma^+$-) transition is a right-hand-circularly polarized (left-hand-circularly polarized) when propagating along the +z-axis (forward), as is well known. When the emission propagates along -z-axis (backward direction), it is left-hand-circularly polarized (right-hand-circularly polarized) [36]. The ratio of $r^L_{+/-}(\sigma^-) = g^{(L)}_{+k} / g^{(L)}_{-k} = \sqrt{\left(\Gamma^{(L)}_{+k} / \Gamma^{(L)}_{-k}\right)}$ can be used for describing the couplings of $\sigma^-$-transition to the forward and backward left-hand-circularly polarized (LHCP) emissions, where $\Gamma^{(L)}_{+k}$ ($\Gamma^{(L)}_{-k}$) denotes the rate of forward (backward) LHCP spontaneous emissions on the $\sigma^-$-transition. In a previous work [36], to precisely measure the ratio, Glan laser prisms, having the highest polarization extinction, were used to distinguish the forward and backward Stokes photons from a ring cavity. In that work, the measured result was $r^L_{+/-}(\sigma^-) \approx 38:1$, showing an excellent nonreciprocal coupling for the atom–photon system.

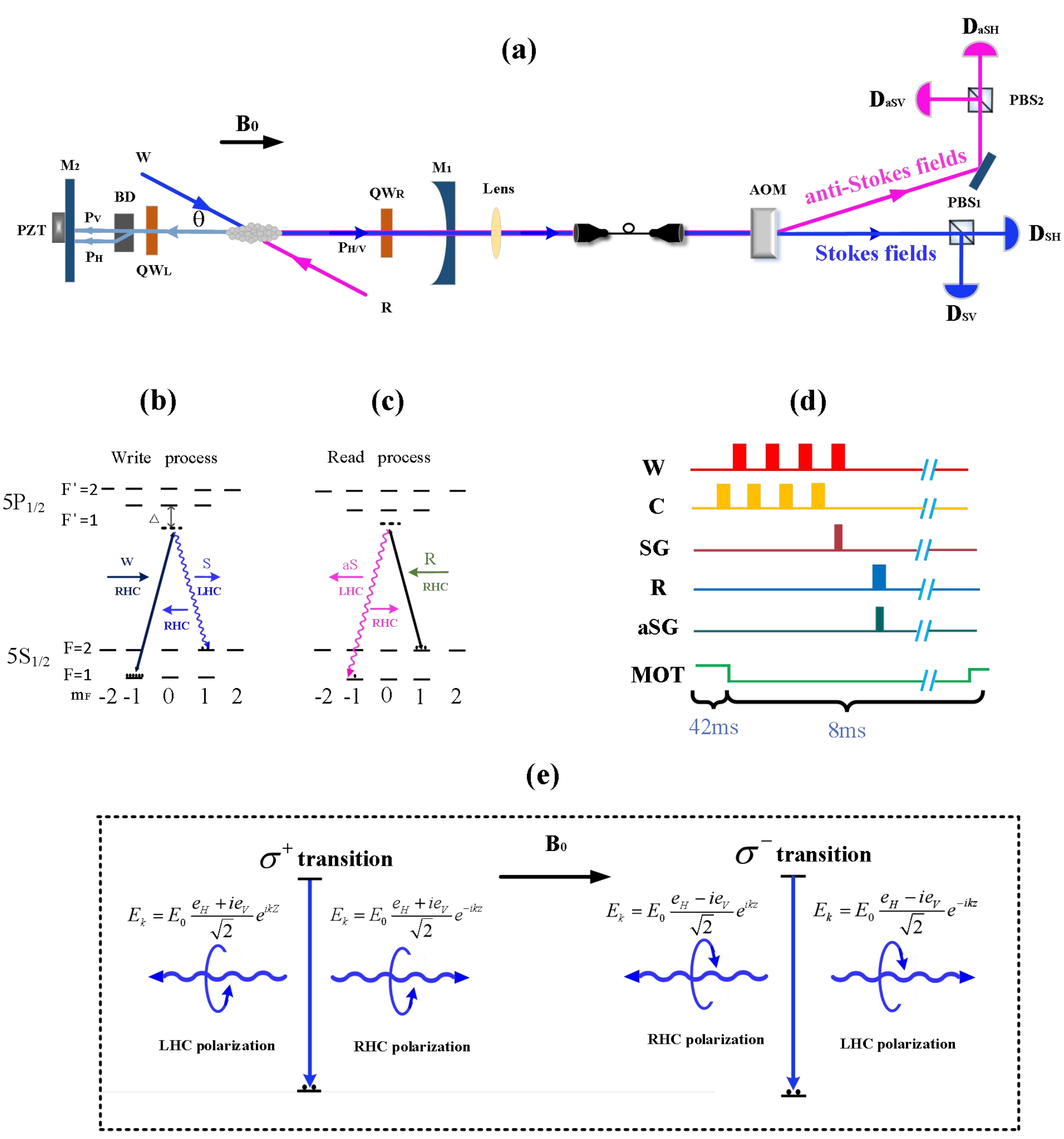

Fig. 1. Overview of the experiment. (a) Setup for the experiment. W: write laser pulse; R: read laser pulse; $M_1$: output coupler mirror (with a reflection of R=96% and a radius of curvature of 400mm) ; $M_2$: high reflectivity ( $R_2$=99% ) flat mirror; PZT: piezoelectric

transducer; PBS: polarizing beam splitter; BD: beam displacer with a thickness of 1 mm; $QW_{L(R)}$: ( $\lambda/4$ ) plate; AOM: acousto–optic modulator; $D_{SH/SV}$, $D_{aSH/aSV}$: single photon detectors. The Stokes and anti-Stokes photons pass through the AOM at different times and are then split. (b) and (c) Relevant atomic levels. The write laser (W) is tuned to $|s_1\rangle = |5S_{1/2}, F=1, m_F=-1\rangle \leftrightarrow |e\rangle = |5P_{1/2}, F'=1, m_F=0\rangle$ with a detuning of $\Delta \approx -108$ MHz. The angle between z-axis and the write (read) beam is $\theta \approx 4.6^\circ$ ($180^\circ - \theta \approx 175.4^\circ$). (d) Time sequence for the experimental trials. W, C, R: write, cleaning, and read laser pulses. S (aS): timeline of the Stokes (anti-Stokes) photon detector gate; MOT: timeline of the magneto–optical trap gate. (e) Diagram showing the polarizations of emission fields from $\sigma^+$ and $\sigma^-$-polarized transitions of the atoms when propagating along the $\pm z$ -axis.

The experiment on spontaneous emissions of atoms in the F–P cavity was then performed. The time sequence of the experimental trials is shown in Fig. 1d. After the $^{87}$Rb atoms were released from MOT, the bias $B_0$ field was applied. Via optical pumping, the atoms were initially prepared into the Zeeman state $|g\rangle = |5S_{1/2}, F=1, m_F=-1\rangle$. Next, a 795-nm "write" laser pulse with a duration of ~100 ns was applied to the atoms at a small angle $\theta$ relative to the z-axis, passing through the center of the atoms. The "write" pulse is right-hand-circularly polarized (RHCP) and interacts with the $|g\rangle \leftrightarrow |e\rangle$ transition. Consequently, spontaneous Raman scattering on the $\sigma^-$-transition $|e\rangle \leftrightarrow |s\rangle$ or $\sigma^+$-transition $|e\rangle \leftrightarrow |s'\rangle$ is induced, probabilistically producing non-classically correlated pairs of a Stokes photon and a spin wave [48]. In the present experiment, only the Stokes field associated with the $\sigma^-$-transition is resonant with the F–P cavity, while that associated with the $\sigma^+$-transition is far off the cavity due to Zeeman states lifting. The Stokes fields initially emitted from the $\sigma^-$-transition, named as $S_{\sigma-}$, are LHCP when propagating along +z (forward) and RHCP along –z (backward) axis. The spin waves correlated with LHCP (RHCP) Stokes photon $S_{\sigma-}$ are denoted as $\mathrm{SW}_+^L$ ($\mathrm{SW}_-^R$), which are defined by the wave vector $\Delta\vec{k}_+ = \vec{k}_w - \vec{k}_{S+}$ ($\Delta\vec{k}_- = \vec{k}_w - \vec{k}_{S-}$), where $\vec{k}_w$, $\vec{k}_{S+}$, and $\vec{k}_{S-}$ are the wave vectors of the write laser and the forward and backward Stokes fields $S_{\sigma-}$, respectively.

At first, the case in which the Stokes field $S_{\sigma-}$ is coupled to a conventional F–P cavity without $\lambda/4$ plates and a BD is considered. The Stokes field $S_{\sigma-}$ propagating along the +z-axis (-z-axis) is LHCP (RHCP) and can be written as $E_+^L(t=0) = E_0 \frac{e_H - ie_V}{\sqrt{2}} e^{-ikz}$ ($E_-^R(t=0) = E_0 \frac{e_H - ie_V}{\sqrt{2}} e^{ikz}$). It is reflected by mirror MR1 (MR2) and then propagates along the –z (+z) axis. The reflected field is written as $E_-^R(\mathrm{Ref_1}) \propto E_0 \frac{e_H - ie_V}{\sqrt{2}} e^{ikz}$ ($E_+^L(\mathrm{Ref_1}) \propto E_0 \frac{e_H - ie_V}{\sqrt{2}} e^{-ikz}$), which is RHCP (LHCP) [45]. It interacts with the $|s\rangle \leftrightarrow |e\rangle$ transition again when passing through the atoms. Therefore, the atom–photon interactions are reciprocal in a conventional F–P cavity. The intracavity Stokes field $E_C$ is a superposition of the forward (+z) and backward (-z) Stokes fields $E_+^L$ and $E_-^R$, which can be written as a standing-wave field $E_C = E_C^0 \cos kz$, where $E_C^0 \propto E_0 \frac{e_H - ie_V}{\sqrt{2}}$. It is noted that spin-wave storage has been demonstrated in a conventional F–P cavity in [2], where the storage lifetime is only up to 0.5 μs due to this superposition.

In the proposed chiral F–P cavity, the Stokes field $S_{\sigma-}$ propagating along the z-axis ($E^L(t=0)=E_0\frac{e_H-ie_V}{\sqrt{2}}e^{-ikz}$) passes through $(\lambda/4)_R$, which converts it from LHCP to V polarization (see Fig. 1S(a) of Supplementary Information). The V-polarized Stokes field is then reflected by mirror $M_1$. The reflected Stokes field again passes through $(\lambda/4)_R$, which converts its polarization from V polarization back to LHCP. After $(\lambda/4)_R$, the Stokes field is expressed as $E^L_{-k}(Ref_1)=E_0\frac{e_H+ie_V}{\sqrt{2}}e^{ikz}$. The reflected Stokes field $E^L_{-k}(Ref_1)$ does not interact with the atomic transition $|e\rangle\leftrightarrow|s\rangle$ when passing through the atoms along the backward direction (–z axis) (see Fig. 1e). This interacts with the $|e\rangle\leftrightarrow|s'\rangle$ transition. However, due to the lifting of Zeeman sublevels by the bias $B_0$, the frequency of the transition $|e\rangle\leftrightarrow|s'\rangle$ is far from the cavity resonance (where the cavity linewidth is ~4 MHz). Thus, coupling of the reflected Stokes field $E^L_{-k}(Ref_1)$ with the $|e\rangle\leftrightarrow|s'\rangle$ transition is negligible. Subsequently, the Stokes field $E^L_{-k}(Ref_1)$ passes through $(\lambda/4)_L$, and its polarization changes from LHCP to V-polarized. The V-polarized Stokes field passes through the BD along the $P_V$ path and is reflected by mirror $M_2$. The Stokes field reflected by $M_2$ then goes through $(\lambda/4)_L$ along the z-axis, whose polarization changes back to LHCP. When this LHCP Stokes field passes through the atoms, it interacts with the atomic transition $|e\rangle\leftrightarrow|s\rangle$. Based on the above process, the forward Stokes field $S_{\sigma-}$ interacts with the atoms only when propagating along the z-axis.

When the Stokes field $S_{\sigma-}$ is initially emitted along the –z-axis, it is RHCP and is written as $E^{R}(t=0)=E_0\frac{e_H-ie_V}{\sqrt{2}}e^{ikz}$. Similar to the interaction of the forward Stokes field $S_{\sigma-}$ with the atoms (Fig. 1S(b) of Supplementary Information), the RHCP Stokes field interacts with the atoms only when it propagates along the –z-axis. Therefore, in the proposed chiral F–P cavity, the LHCP or RHCP of the light field is preserved, and the interactions of the Stokes fields $S_{\sigma-}$ with the atoms become nonreciprocal.

Nonreciprocal atom–photon couplings require the chiral cavity to be individually resonant with the LHCP or RHCP Stokes fields. This is achieved by controlling the temperature of the BD, where its refraction for H-polarized- and V-polarized light varies with temperature. When the LHCP (RHCP) Stokes field is resonant with the cavity length, the atoms experience it as a traveling-fields: $E_C^L=E_C\frac{e_H-ie_V}{\sqrt{2}}e^{-ikz}$ ($E_C^R=E_C\frac{e_H-ie_V}{\sqrt{2}}e^{ikz}$), where $E_C=E_0\sqrt{F_0/2\pi}$ denotes the intracavity field, and $F_0$ denotes the cavity finesse, whose value is ~50 in this experiment. The intracavity Stokes fields escape through the mirror $M_1$. The escaped LHCP (RHCP) mode is a V-polarized (H-polarized) field, written as $E^{L+}(T)\propto\left(F_0/2\pi\right)E_0\vec{e}_V e^{-ikz}$ ($E^{R-}(T)\propto\left(F_0/2\pi\right)E_0\vec{e}_H e^{-ikz}$), where $T\approx 4\%$ denotes the transmission of $M_1$. As shown in Figure 1a, the output V-polarized (H-polarized) Stokes photons pass through the AOM and enter $PBS_1$, which reflects V-polarized and transmits H-polarized photon. The LHCP (RHCP) Stokes field is non-classically correlated with the spin wave $SW_L^+$ and $SW_R^-$, meaning that a photon detection event in the $E^L(T)$ ($E^R(T)$) field heralds the storage of $SW_L^+$ ($SW_R^-$). As shown in Figure 1a, the output Stokes fields $E_C^L$ and $E_C^R$ were detected by single-photon detectors $D_{SV}$ and $D_{SH}$, respectively. A detection event at $D_{SV}$ ($D_{SH}$) heralds the storage of $SW_L^+$ ($SW_R^-$).

The spin waves $SW_L^+$ and $SW_R^-$ can be efficiently converted into anti-Stokes photons by applying a read laser pulse propagating opposite to the write beam. Under the phase-matching condition [36], the anti-Stokes field retrieved from $SW_L^+$ ($SW_R^-$) propagating along –z (+z) is LHCP (RHCP). The field retrieved from $SW_L^+$ ($SW_R^-$) resonates with the cavity and is enhanced [48]; its output from $MR_1$ is V-polarized (H-polarized) and goes through the AOM. The AOM routes the V-polarized and H-polarized anti-Stokes fields to $PBS_2$. The former is then directed into the single-photon detector $D_{aSV}$, and the latter into $D_{aSH}$.

Because the spin waves $SW_L^+$ and $SW_R^-$ are non-classically correlated with the forward and backward Stokes emissions, respectively, the retrieval efficiencies from $SW_L^+$ and $SW_R^-$ are defined by $\gamma_L^+(t) = P_{S,aS}^L / \eta P_S^L$ and $\gamma_R^-(t) = P_{S,aS}^R / \eta P_S^R$ [5], where $P_{S,aS}^L$ denotes the coincidence probability between the LHCP Stokes and anti-Stokes photons, $P_{S,aS}^R$ denotes the coincidence probability between the RHCP Stokes and anti-Stokes photons, $P_S^L$ ($P_S^R$) denotes the probability of detecting a Stokes photon at $D_{SV}$ ($D_{SH}$), and $\eta$ denotes the detection efficiencies of the anti-Stokes channels. Considering the decoherence of spin waves, the retrieval efficiencies $\gamma_{L,R}^{\pm}(t)$ are rewritten as [5, 50]:

$$\gamma_L^+(t) = \gamma_0 \left[1+\left(vt/r_0\right)^2\right]^{-1} e^{-(t/\tau_+)^2} \quad (1a)$$

$$\gamma_R^-(t) = \gamma_0 \left[1+\left(vt/r_0\right)^2\right]^{-1} e^{-(t/\tau_-)^2} \quad (1b)$$

where, $\gamma_0$ is the retrieval efficiency at $t = 0$, $\tau_{\pm} = 1/\left|\Delta\vec{k}_{\pm}\right|\bar{v}$ denotes the decoherence time of the spin waves caused by atomic motions [2]; $\left|\Delta\vec{k}_+\right| = \left|\vec{k}_w - \vec{k}_{S+}\right| \approx 2k_0 \sin\theta/2$, $\left|\Delta\vec{k}_-\right| = \left|\vec{k}_w - \vec{k}_{S-}\right| \approx 2k_0 \cos\theta/2$ denote the absolute values of wave vectors $SW_L^+$ and $SW_R^-$, respectively; $\bar{v} = \sqrt{k_B T/m}$ is the average velocity of the atoms; $\left[1+\left(vt/r_0\right)^2\right]^{-1}$ denotes the expanding factor of the atoms, and $r_0$ is the radius of the Stokes field in the cavity. Because the decoherence rate of the spin wave $SW_R^-$ is significantly higher than that of the spin wave $SW_L^+$, the former has a shorter lifetime than the latter. Therefore, the dependences of the retrieval efficiencies for the spin waves $\gamma_L^+(t)$ and $\gamma_R^-(t)$, which are correlated with the LHCP and RHCP Stokes fields, respectively, on the storage time $t$ should be split into two curves, in contrast to the single curve observed in the conventional F–P cavity. By observing these two dependences, it can be determined whether the atom–photon coupling in the F–P chiral cavity is nonreciprocal.

## III. EXPERIMENTAL RESULTS.

The free spectral range of the LHCP and RHCP Stokes fields is 214 MHz. In the measurement of $\gamma_L^+(t)$ ($\gamma_R^-(t)$), the F–P cavity is tuned to be resonant with the LHCP (RHCP) Stokes field, whereas the RHCP (LHCP) field is detuned by 107 MHz.

The triangles in Fig. 2 represent the measured time-dependent retrieval efficiencies of the spin waves correlated with the LHCP Stokes fields, and the square dots represent those of the spin waves correlated with the RHCP Stokes fields. The red curve (a) in Fig. 2 fits the triangle-dot data according to $\gamma_L^+(t)$ in Equation (1a), and the blue curve (b) is the fits to $\gamma_R^-(t)$ in Equation (1b). The fits according to Equation (1a) and (1b) align with the measured data, which shows that the spontaneous emission fields along forward and backward directions in the chiral F–P cavity are split into two traveling-wave modes.

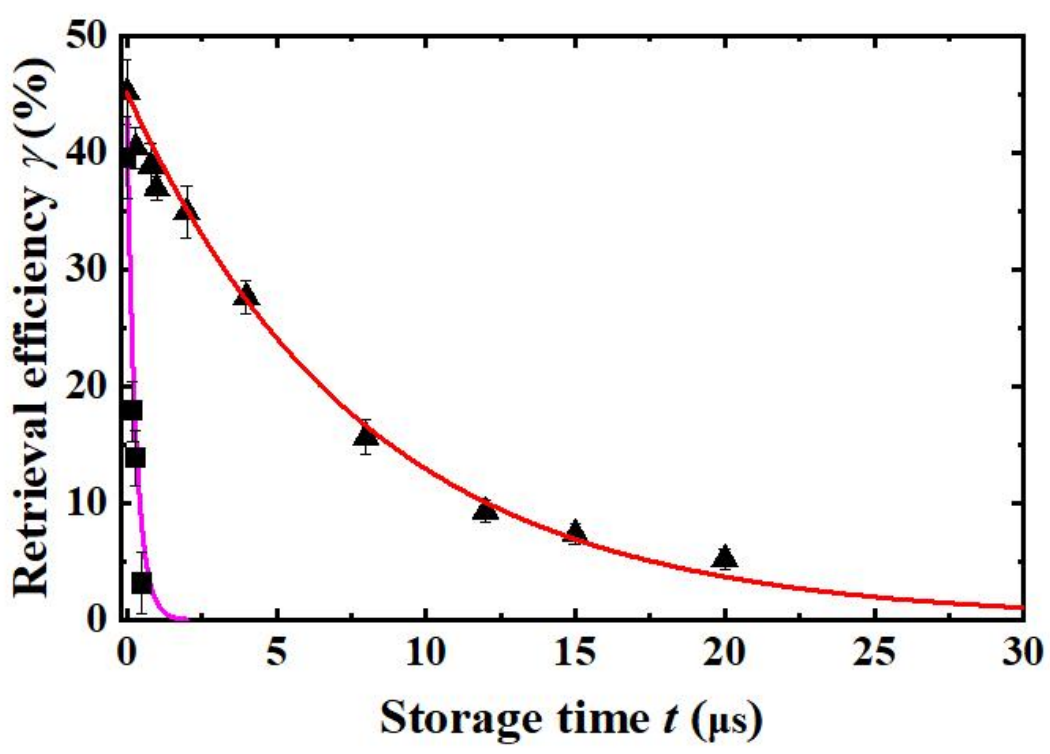

Figure 2. Measured time-dependent retrieval efficiencies of spin waves correlated with LHCP Stokes fields (triangle dots) and the retrieval efficiencies of spin waves correlated with RHCP Stokes fields (square dots). The red and blue curves are the fits to the measured triangle and square-dot data according to the $\gamma_L^+(t)$ and $\gamma_R^-(t)$ in Equation (1a) and (1b), respectively, which yield a longer (8μs) and a shorter (300ns) lifetime, respectively. The fitting parameters $\theta \approx 4.6^\circ$, $T \approx 100\mu k$, $r_0 \approx 300\mu m$ are derived from experimental values.

The photon count ratio of the V- to H-polarized Stokes fields, as defined by $k = P_S^L / P_S^R$, was also measured. When the LHCP (RHCP) Stokes field was resonant with the cavity, the ratio was $k \approx 20:1$ ($k \approx 1:20$), indicating effective splitting of the LHCP and RHCP fields in the measurements of $\gamma_L^+(t)$ and $\gamma_R^-(t)$.

## IV. CONCLUSION.

The nonreciprocal couplings of the spontaneous-emission fields from a single circularly polarized ($\sigma^-$) transition to the atoms in a chiral F–P cavity was experimentally demonstrated. In contrast to the result obtained in a conventional F–P cavity [2], where, the intracavity Stokes field is a superposition of forward and backward Stokes fields, the intracavity field in the proposed chiral F–P cavity is split into two traveling-wave modes: LHCP and RHCP. By distinguishing the two modes, long-lifetime storage in the F–P cavity is achieved, which provides a feasible approach for realizing efficient and long-lifetime quantum storage in integrated quantum memories based on micro atomic ensembles coupled to fiber-based F–P cavities. In the present experiment, the long-lived memory lifetime is limited to 8μs due to a larger angle $\theta$. However, this is not a fundamental limitation, and the lifetime can be significantly increased by decreasing the angle $\theta$ [48]. It is further shown that achieving nonreciprocal (traveling-wave) atom–light coupling in an F–P cavity requires that the light field be emitted from a circularly polarized atomic transition, that the F–P cavity coupled with the field be chiral, and that the atom–light system breaks the TRS. The proposed chiral atom–cavity scheme provides a road to achieve spatially uniform couplings between single atoms and an F–P cavity, which is beneficial for the development of effective cavity-array microscopes for parallel single-atom interfacing [51, 52].

**DISCLOSURES.**

The authors declare no conflicts of interest.

**DATA AVAILABILITY.**

Data may be obtained from the authors upon reasonable request.

**ACKNOWLEDGMENTS.**

The Fund for Quantum Science and Technology-National Science and Technology Major Project (2025ZD0300200), the Fund for National Natural Science Foundation of China (12574538 and 12174235), the Fund for Shanxi Key Subjects Construction (1331), and the Fundamental Research Program of Shanxi Province (202203021221011)